# NANOMECHANICAL RESONANT STRUCTURES IN SINGLE-CRYSTAL DIAMOND

*Michael J. Burek [a], Daniel Ramos [a], Parth Patel [b], Ian W. Frank [a], and Marko Lončar [a,†]*

[a.] School of Engineering and Applied Sciences, Harvard University, 29 Oxford Street, Cambridge, MA 02138, USA

[b.] University of Waterloo, 200 University Avenue West Waterloo, ON N2L 3G1, Canada

[†] Corresponding author contact: E-mail: loncar@seas.harvard.edu. Tel: (617) 495-579. Fax: (617) 496-6404.

**ABSTRACT** – With its host of outstanding material properties, single-crystal diamond is an attractive material for nanomechanical systems. Here, the mechanical resonance characteristics of freestanding, single-crystal diamond nanobeams fabricated by an angled-etching methodology are reported. Resonance frequencies displayed evidence of significant compressive stress in doubly clamped diamond nanobeams, while cantilever resonance modes followed the expected inverse-length-squared trend. Q-factors on the order of $10^4$ were recorded in high vacuum. Results presented here represent initial groundwork for future diamond-based nanomechanical systems which may be applied in both classical and quantum applications.

The outstanding material properties exhibited by diamond make it an attractive nanomechanics platform, due in part to its high Young's modulus, high thermal conductivity, and low intrinsic dissipation[1]. In addition, diamond is an impressive optical material, with its wide transparency window and vast inventory of luminescent defects[2]. In particular, the negatively charged nitrogen vacancy (NV$^-$), has garnered significant attention from its operation as a room temperature single photon source[3] and an optically addressable solid-state spin-qubit[4]. Integration of NV$^-$ centers with diamond nanomechanical systems would ultimately enable scalable quantum information architectures[5] and sensitive nanoscale magnetometry[6]. Despite its promise, single-crystal diamond nanomechanical resonators have progressed slowly, primarily due to a lack of heteroepitaxial growth techniques. With the emergence of wafer-scale nanocrystalline diamond thin films on supporting substrates[7,8], micron[9] and nanometer[10] scale mechanical systems have been realized in this material. However, their performance is often limited by inferior properties due to grain boundaries, impurities, and large interfacial stresses[7].

To date, fabrication of single-crystal diamond nanomechanical systems has mainly utilized crystal ion slicing[11], where irradiation of a single-crystal diamond substrate with energetic ions creates a sub-surface graphitized layer which can be selectively removed. Though, newly developed homoexpitaxial regrowth on single-crystal diamond membranes released by ion slicing[12] have advanced this fabrication methodology. However, residual crystal damage and large film stresses resulting from ion implantation significantly reduces final device quality[11,13]. Recently, small scale single-crystal diamond cantilevers[14] – exhibiting mechanical Q-factors >$10^5$ – were fabricated via a diamond-on-insulator platform[15], created by thinning diamond slabs adhered to a supporting substrate. Though this approach remains promising, without heteroepitaxially grown thin films, further development of diamond nanomechanical systems requires exploring alternative device fabrication avenues.

Recently, we developed the three-dimensional nanofabrication of suspended nanostructures from bulk single-crystal diamond substrates. This fabrication process, described in detail elsewhere[16], employs anisotropic plasma etching performed at an oblique angle to the substrate surface (referred to hereafter as 'angled-etching'). Angled-etching yields freestanding nanobeams – with triangular cross-sections – directly from single-crystal diamond substrates. In this work, the nanomechanical properties of doubly clamped nanobeam resonators and cantilevers fabricated in single-crystal diamond by angled-etching are presented.

Figure 1 displays SEM images of diamond nanomechanical resonators. Freestanding nanobeams were 10 to 85 μm long, with widths between 500 nm and 1.3 μm. Optical grade, <100>-oriented, single-crystal diamonds (< 1 ppm [N], Element Six) were used throughout this work. In general, angled-etching fabrication offers excellent yield (> 95%) and throughput (> $10^3$ devices/mm$^2$), with over 500 devices measured in this work. Nanobeams shown in Figure 1 (a) and (b), which are doubly clamped nanomechanical resonators, are suspended above the diamond substrate with significant clearance (~ 1.5 μm). As displayed in Figure 1 (c), focused ion beam (FIB) milled cross-sections revealed a symmetric triangular shape, with a bottom apex half-angle of $\theta \sim 50°$. With angled-etching, the nanobeam thickness (*t*) is linked to the prescribed width (*w*) via the relationship $t = w/(2\tan\theta)$. Figure 1 (d) and (e) show images of as-fabricated single-crystal diamond nanobeam cantilevers, which displayed identical substrate clearance and cross-section to doubly clamped nanobeams.

The transverse vibrations of rigid nanobeams are described by Euler-Bernoulli theory[17], which yields a relationship for nanobeam resonance frequency that is inversely proportional the length squared and directly proportional to the width. Since the diamond nanobeam cross-section is not axially symmetric, out-of-plane and in-plane bending moments are $I_x = wt^3/36$ and $I_y = w^3t/48$, respectively. As such, the geometry dependent resonance frequencies, $f_n$, are:

$$f_{x,n} = \frac{1}{(2\pi)} \frac{k_n^2}{l^2} \sqrt{\frac{Et^2}{18\rho}} \tag{1a}$$

and

$$f_{y,n} = \frac{1}{(2\pi)} \frac{k_n^2}{l^2} \sqrt{\frac{Ew^2}{24\rho}} \tag{1b}$$

where $E$ is the Young' modulus, $\rho$ is the material density, and $l$ is the nanobeam length. The parameter $k_n$ is a mode index, which is determined from boundary conditions associated with the specific type of nanobeam resonator[17]. From the triangular nanobeam cross-section in Figure 1 (c), Eq. (1) indicates the same order out-of-plane vibrations occur at lower frequencies than in-plane ones.

Diamond nanobeam Brownian motion was characterized via optical interferometric displacement detection, employing a focused laser at normal incidence to the substrate[18]. In this detection scheme, interference between light confined within the nanobeam and the standing light wave – which originates from the incident laser interfering with light reflected from the substrate – provided ample displacement sensitivity to out-of-plane motion. Additionally, optical characterization of nanobeam resonators in a similar configuration has previously been shown to provide comparable sensitivity to in-plane resonator motion[19]. Details regarding the detection limits afforded by this technique are discussed later in the text. The characterization set up, shown schematically in Figure 1 (f), employed a tunable telecom laser focused by an objective (NA = 0.5), through a vacuum chamber view port, on the diamond nanobeams. Individual nanobeams were positioned under the focused laser spot using motorized stages. Light reflected from the sample was collected by a photodetector (New Focus model 1811). A spectrum analyzer at the photodetector output revealed mechanical resonances. All measurements were conducted at $\sim 10^{-6}$ Torr and room temperature.

Figure 2 (a) displays a thermomechanical power spectral density (PSD) collected from a 675 nm wide, 55 μm long doubly clamped diamond nanobeam. High-resolution thermal noise spectra of the observed resonances are displayed as insets. Since diamond nanobeams are not externally actuated, the voltage spectral density of the thermal fluctuations, $S_v^{1/2}$ (μV/Hz$^{1/2}$), may be converted to a displacement spectral density, $S_z^{1/2}$ (pm/Hz$^{1/2}$), via equipartition theorem[20]. To do so, the spring constant of each mode was determined via finite element method (FEM) simulations (COMSOL Multiphysics)[21]. Fitting to calibrated thermomechanical spectra gave the resonance frequency, mechanical Q-factor, conversion from voltage to displacement, and displacement noise floor. The mechanical Q-factors are approximately 9,400, 10,900, and 19,000 for the three mechanical modes, respectively.

Mechanical resonance frequencies measured from ~ 565 and 675 nm wide doubly clamped nanobeams are plotted in Figure 2 (b) and (c) as a function of length. Here, only the first three observed resonance frequencies are plotted. Although the higher-frequency resonances appear to follow an inverse power law relationship with length, the lowest frequency mode displays a complicated behavior, not following trends predicted by Euler-Bernoulli theory. Such a discrepancy between theory and experiment is a strong indication the diamond nanobeams are compressively stressed[22-23], which modifies the expected resonance frequencies as:

$$f_n(\sigma) = f_n \sqrt{1 - \frac{\sigma A l^2}{k_n^2 EI}} \tag{2}$$

where $\sigma$ is the magnitude of uniaxial compressive stress along the length of the beam and $A$ is the cross-sectional area. As such, compressive stress lowers the resonance frequency, with the term under the square root in Eq. (2) vanishing when the stress approaches the critical Euler buckling load $\sigma_c$, which for a doubly clamped beam, is defined as:

$$\sigma_c = \frac{\pi^2 EI}{(0.5l)^2 A} \tag{3}$$

In the context of Figure 2 (b) and (c), Eqs. (2) and (3) may be alternatively viewed in terms of a critical buckling length, $l_c$, given a built-in compressive stress. Previous reports noted that experimental resonance frequencies of polycrystalline silicon resonators were strikingly different for beams longer than the critical-buckling length dictated by a built-in compressive film stress[23], while shorter nanobeams were well approximated by Eq. (2). Similar observations are made here in Figure 2 (b), where the first experimental resonance mode of 565 nm wide nanobeams reaches a minimum near $l \sim 25$ µm, beyond which the resonance frequencies increase and level off until $l \sim 40$ µm. Away from this point, resonance frequencies eventually recover an inverse-power-law relationship. The local minimum near $l \sim 25$ µm represents the transition from a compressively-stressed, unbuckled nanobeam to a buckled nanobeam (i.e. critical buckling length). The increase in resonance frequency for lengths slightly greater than 25 μm is likely due to the significant release of compressive stress through buckling deformation, while nanobeams much longer than 25 μm will recover the frequency-versus-length trend predicted by Euler-Bernoulli theory. A similar trend, though less pronounced, is observed in Figure 2 (c) for 675 nm wide diamond nanobeams, with a minimum in the first experimental resonance now near $l \sim$ 40 μm.

FEM simulations were employed to fit the experimental data in Figure 2 (b) and (c), as shown by dashed lines. Built-in compressive stress was applied by gradually increasing the initial material strain. By employing non-linear solvers and this monotonically increasing parameter, the software is able to avoid the bifurcation in solutions normally associated with the buckling of a rigid beam and calculate the final deformed shape for a given compressive stress. A non-linear eigenfrequency solver is then used to calculate the mechanical modes of the resulting stressed and deformed structures. Their simulated flexural shape allowed discrimination between in-plane and out-of-plane vibrations, as well as determining the mode order by the number of anti-nodes. The simulations show excellent agreement with experimental data for all three plotted modes, and interestingly, also predict a spectral crossing of the first and second resonance mode shapes. It is important to note the fitting parameters employed in the simulations (initial material strain and Young's modulus) were extremely sensitive to the experimental data near and to the left of the buckling transitions.

From simulation, the estimated built-in compressive stress was ~ 140 MPa, assuming a diamond Young's modulus of roughly 900 GPa. Origins of compressive stress in fabricated nanobeams are not entirely clear, especially since the nanostructures are fabricated from a bulk crystal, making interfacial stresses unlikely. Presumably, stress may have originated from the etch mask used during fabrication, though further investigation of stress in diamond nanobeams is beyond the scope of the present study. Knowledge of built-in stress in angled-etched nanobeams – in the context of diamond nanomechanics with integrated NV⁻ or other color centers – is particularly important since spectral properties of diamond color centers are impacted by both local and global lattice perturbations[2].

Ultimately, the potentially adverse effects of compressive stress on resonance frequency and mechanical Q-factor are circumvented in nanobeam cantilevers, where axial stress is released by the free-ended structure. Figure 3 (a) displays a thermomechanical power spectral density for an 880 nm wide, 20 µm long cantilever. Two resonance peaks are revealed, with lower and higher frequencies now – by Euler-Bernoulli theory – attributed to out-of-plane and in-plane flexural modes, respectively. By Eq. (1), the ratio of the same order in-plane to out-of-plane resonance frequencies reflects the etch angle through $\theta = \tan^{-1}\left(f_{y,n}/\sqrt{3}f_{x,n}\right)$. Applying this relation to the experimental data gave an etch angle of 53.5º ± 2º, which is an ensemble estimate and in close agreement with that obtained from FIB cross-sections. Measured out-of-plane cantilever resonances are plotted in Figure 3 (b). Here, the expected $f \propto l^{-2}$ trend is clear. Dashed lines in Figure 4 (b) are calculated with Eq. (1a), using appropriate beam geometries and $\rho$ = 3500 kg/m³. From the fits, the Young's modulus was estimated to be ~ 901 ± 58 GPa. This low modulus value for single-crystal diamond is likely due to the high level of nitrogen doping in the substrates[24]. We note that diamond nanomechanics from single-crystal substrates developed here enables investigation of resulting material properties and processes latitude for synthetic diamond growth in a chip-scale manner, as has previously been done with silicon-based substrates and thin films[23,25].

High-resolution thermal noise spectra of the out-of-plane and in-plane resonance peaks displayed in Figure 3 (a) are shown in Figures 3 (c) and (d), respectively. Again, thermomechanical calibration is carried out on the acquired spectra. The mechanical Q-factors, estimated from FWHM of the peaks, are approximately 47,800 and 50,800 for the two modes, respectively. The highest measured Q-factor in the range of fabricated cantilevers was ~ 94,000. Q-factors estimated for fundamental out-of-plane diamond cantilevers resonances are plotted in Figure 3 (e), with the dashed line as a guide for the eye. From Figure 3 (e), the mechanical dissipation displays a limited dependence on length, though higher Q-factors for longer nanobeams is apparent. This likely suggests diamond cantilevers are limited by clamping losses, and longer devices would increase mechanical Q-factor. However, further study of dissipation mechanisms in diamond resonators is beyond the scope of the current study.

Future applications of diamond nanomechanics (i.e. spin optomechanics with $NV^-$ centers[26]) will ultimately require highly sensitive displacement detection. Thus, it is important to elucidate the mechanism responsible for optical characterization of nanobeam motion in this work. Since the diamond nanobeam widths are on-the-order-of or smaller than the telecom wavelength probe laser, and the fact in-plane cantilever motion is readily observed, more than simple scattering of light is required to detect the levels of movement measured. Intuitively, implementing a visible wavelength laser would enhance interferometric displacement detection, and indeed, the majority of prior works employ a red laser for optical interferometry measurements of nanomechanical systems[18,19-20]. However, this would likely complicate visible spectroscopy on $NV^-$ centers, while also optically probing mechanical motion. Therefore, we offer insights on how optical displacement detection may be optimized as a function of nanobeam cross-sectional geometry, which is especially prevalent for angled-etched nanobeams since device width, thickness, and distance above the substrate may be engineered via the etch time[16].

Figure 4 (a) and (b) show FEM (COMSOL) simulations of 350 and 1000 nm wide diamond nanobeams in a focused Gaussian laser beam ($\lambda$ = 1500 nm, incident power $P_o$ = 100 µW). The spacing between the nanobeam bottom apex and substrate (hereinafter referred as 'substrate distance') was fixed at 1.0 µm, while nanobeams were laterally centered on the optical axis. A spatial confinement of the electric field intensity within the nanobeam cross-section is observed. In other words, diamond nanomechanical structures also support optical resonances – so called leaky modes[27,28] – arising from light trapped by multiple internal reflections. Therefore, laser photons focused on a diamond nanobeam may take one of two paths: i) they are reflected by the substrate and interfere with the incident laser beam resulting in a standing wave, or ii) they are trapped inside the nanobeam by exciting its optical resonance and then re-emitted. The latter – known as resonant scattering – has been used to characterize various nanophotonic devices[29]. Nanobeam motion modulates the overlap between the leaky optical resonance and stationary standing wave, resulting in ample displacement sensitivity for both in-plane and out of plane mechanical vibrations. Use of similar confined electromagnetic modes to extend optomechanics to sub-wavelength silicon nanowires was recently reported[28]. Far field profiles of backscattered laser intensity were generated for an experimentally relevant range of widths, substrate distances, and lateral displacements from the optical axis, with an example shown in Figure 4 (c). From this figure, backscattered light from a 250 nm wide nanobeam is largely due to light reflected by the substrate. However, with increasing width, emerging side lobes significantly modified the far field profile, which is attributed to interference between optical modes confined by the nanobeam and the standing light wave.

To further quantify optomechanical read-out of nanobeam motion, reflectivity values were calculated by integrating the backscattered light intensity over the objective lens' numerical aperture (i.e. +/- 30°) for the series of simulated far field profiles and normalized by the incident power, in a similar manner to Karabacak *et al.*[19]. The system responsivity (units of A/m) to nanobeam displacements was then extracted given the photodetector responsivity (~ 1 A/W at $\lambda$ = 1500 nm) and the spatial gradient of the reflectivity for a given nanobeam width as a function of either the range of substrate distances (responsivity for out-of-plane motion) or nanobeam lateral displacements (responsivity for in-plane motion). Finally, the simulated system responsivity was used to convert the experimental noise floor, set by the photodetector current noise spectral density, $S_I^{1/2}$ (~ 2 nA/Hz$^{1/2}$ over the range of measured nanobeam mechanical resonances), to a displacement noise spectral density, $S_z^{1/2}$ (in units of m/Hz$^{1/2}$). The results are represented by color maps shown in Figure 4 (d) and (e) for out-of-plane and in-plane motion, respectively. The simulated displacement noise for out-of-plane motion is in good agreement with the experimental thermomechanical noise spectra displayed previously, where an average displacement noise floor of ~ 0.66 pm/Hz$^{1/2}$ was observed. From these 2D calculations, the optimal nanobeam width and position, which maximizes the sensitivity to nanobeam displacements, may be extracted.

In summary, the mechanical resonance characteristics of single-crystal diamond doubly clamped nanobeams and cantilevers are reported. Resonance frequencies displayed evidence of significant compressive stress in doubly clamped nanobeams, while cantilever resonance modes followed the expected inverse-length-squared trend. Q-factors on the order of $10^4$ were recorded in high vacuum. Results presented here represent initial groundwork for future diamond-based nanomechanical systems which may be applied in both classical and quantum applications.

This work was supported in part by the Defense Advanced Research Projects Agency (QuASAR program), and AFOSR MURI (grant FA9550-12-1-0025). Fabrication was performed at the Center for Nanoscale Systems (CNS) at Harvard University. M.J. Burek is supported in part by the Natural Science and Engineering Council (NSERC) of Canada. D. Ramos acknowledges financial support from the EU Marie Curie grant IOF-2009-254996. The authors thank H. Atikian, A. Maygar, and T.Y. Tsui for valuable discussions.

# REFERENCES

1. S. E. Coe and R. S. Sussmann, Diamond and Related Materials **9** (9–10), 1726 (2000).
2. Igor Aharonovich, Andrew D. Greentree, and Steven Prawer, Nature Photonics **5** (7), 397 (2011).
3. T.M. Babinec, B.J.M. Hausmann, M. Khan, Y. Zhang, J.R. Maze, P.R. Hemmer, and M. Loncar, Nature Nanotechnology **5** (3), 195 (2010).
4. M. V. Gurudev Dutt, L. Childress, L. Jiang, E. Togan, J. Maze, F. Jelezko, A. S. Zibrov, P. R. Hemmer, and M. D. Lukin, Science **316** (5829), 1312 (2007).
5. P. Rabl, S. J. Kolkowitz, F. H. L. Koppens, J. G. E. Harris, P. Zoller, and M. D. Lukin, Nat Phys **6** (8), 602 (2010).
6. MaletinskyP, HongS, M. S. Grinolds, HausmannB, M. D. Lukin, R. L. Walsworth, LoncarM, and YacobyA, Nat Nano **7** (5), 320 (2012).
7. Anirudha V. Sumant, Orlando Auciello, Robert W. Carpick, Sudarsan Srinivasan, and James E. Butler, MRS Bulletin **35** (04), 281 (2010); Orlando Auciello, S. Pacheco, Anirudha V. Sumant, C. Gudeman, S. Sampath, A. Datta, Robert W. Carpick, Vivekananda P. Adiga, P. Zurcher, Ma Zhenqiang, Hao-Chih Yuan, J. A. Carlisle, B. Kabius, J. Hiller, and S. Srinivasan, Microwave Magazine, IEEE **8** (6), 61 (2007).
8. Auciello Orlando, Birrell James, A. Carlisle John, E. Gerbi Jennifer, Xiao Xingcheng, Peng Bei, and D. Espinosa Horacio, Journal of Physics: Condensed Matter **16** (16), R539 (2004); J. K. Luo, Y. Q. Fu, H. R. Le, J. A. Williams, S. M. Spearing, and W. I. Milne, Journal of Micromechanics and Microengineering **17** (7), S147 (2007); Orlando Auciello and Anirudha V. Sumant, Diamond and Related Materials **19** (7–9), 699 (2010).
9. E. Kohn, M. Adamschik, P. Schmid, S. Ertl, and A. Flöter, Diamond and Related Materials **10** (9–10), 1684 (2001); Hadi Najar, Amir Heidari, Mei-Lin Chan, Hseuh-An Yang, Liwei Lin, David G. Cahill, and David A. Horsley, Applied Physics Letters **102** (7), 071901 (2013); Matthias Imboden, Pritiraj Mohanty, Alexei Gaidarzhy, Janet Rankin, and Brian W. Sheldon, Applied Physics Letters **90** (17), 173502 (2007); V. P. Adiga, A. V. Sumant, S. Suresh, C. Gudeman, O. Auciello, J. A. Carlisle, and R. W. Carpick, Physical Review B **79** (24), 245403 (2009).
10. L. Sekaric, J. M. Parpia, H. G. Craighead, T. Feygelson, B. H. Houston, and J. E. Butler, Applied Physics Letters **81** (23), 4455 (2002); A. B. Hutchinson, P. A. Truitt, K. C. Schwab, L. Sekaric, J. M. Parpia, H. G. Craighead, and J. E. Butler, Applied Physics Letters **84** (6), 972 (2004); Alexei Gaidarzhy, Matthias Imboden, Pritiraj Mohanty, Janet Rankin, and Brian W. Sheldon, Applied Physics Letters **91** (20), 203503 (2007); Patrik Rath, Svetlana Khasminskaya, Christoph Nebel, Christoph Wild, and Wolfram H.P. Pernice, Nature Communications **4** (2013).
11. Meiyong Liao, Shunichi Hishita, Eiichiro Watanabe, Satoshi Koizumi, and Yasuo Koide, Advanced Materials **22** (47), 5393 (2010); Maxim K. Zalalutdinov, Matthew P. Ray, Douglas M. Photiadis, Jeremy T. Robinson, Jeffrey W. Baldwin, James E. Butler, Tatyana I. Feygelson, Bradford B. Pate, and Brian H. Houston, Nano Letters **11** (10), 4304 (2011).
12. Igor Aharonovich, Jonathan C. Lee, Andrew P. Magyar, Bob B. Buckley, Christopher G. Yale, David D. Awschalom, and Evelyn L. Hu, Advanced Materials **24** (10), OP54 (2012).
13. Barbara A. Fairchild, Paolo Olivero, Sergey Rubanov, Andrew D. Greentree, Felix Waldermann, Robert A. Taylor, Ian Walmsley, Jason M. Smith, Shane Huntington, Brant C. Gibson, David N. Jamieson, and Steven Prawer, Advanced Materials **20** (24), 4793 (2008); C. F. Wang, E. L. Hu,

| | J. Yang, and J. E. Butler, Journal of Vacuum Science & Technology B: Microelectronics and Nanometer Structures **25** (3), 730 (2007). |
|---|---|
| 14 | P. Ovartchaiyapong, L. M. A. Pascal, B. A. Myers, P. Lauria, and A. C. Bleszynski Jayich, Applied Physics Letters **101** (16), 163505 (2012); Y. Tao, J.M. Boss, B.A. Moores, and C.L. Degen, arXiv:1212.1347v1 (2012). |
| 15 | Andrei Faraon, Paul E. Barclay, Charles Santori, Kai-Mei C. Fu, and Raymond G. Beausoleil, Nature Photonics **5** (5), 301 (2011); Birgit J. M. Hausmann, Brendan Shields, Qimin Quan, Patrick Maletinsky, Murray McCutcheon, Jennifer T. Choy, Tom M. Babinec, Alexander Kubanek, Amir Yacoby, Mikhail D. Lukin, and Marko Loncar, Nano Letters **12**, 1578−1582 (2012); B. J. M. Hausmann, I. B. Bulu, P. B. Deotare, M. McCutcheon, V. Venkataraman, M. L. Markham, D. J. Twitchen, and M. Lončar, Nano Letters **13** (5), 1898 (2013). |
| 16 | Michael J. Burek, Nathalie P. de Leon, Brendan J. Sheilds, B.J.M. Hausmann, Yiwen Chu, Qimin Quan, Alexander S. Zibrov, Hongkun Park, Mikhail. D. Lukin, and Marko Loncar, Nano Letters **12** (12), 6084 (2012). |
| 17 | William Weaver, Stephen P. Timoshenko, and Donovan H. Young, *Vibration problems in engineering*, 5th ed. (Wiley, 1990). |
| 18 | D. W. Carr, L. Sekaric, and H. G. Craighead, Papers from the 41st international conference on electron, ion, and photon beam technology and nanofabrication **16**, 3821 (1998); T. Kouh, D. Karabacak, D. H. Kim, and K. L. Ekinci, Applied Physics Letters **86** (1), 013106 (2005). |
| 19 | D. Karabacak, T. Kouh, C. C. Huang, and K. L. Ekinci, Applied Physics Letters **88** (19), 193122 (2006). |
| 20 | W. K. Hiebert, D. Vick, V. Sauer, and M. R. Freeman, Journal of Micromechanics and Microengineering **20** (11), 115038 (2010). |
| 21 | Thermomechanical calibration of doubly clamped diamond nanobeams took into account the built up compressive stress determined through FEM fits to the frequency versus length plots in Figure 2 (b) and (c). . |
| 22 | Jun Seong Chan, X. M. H. Huang, M. Manolidis, C. A. Zorman, M. Mehregany, and J. Hone, Nanotechnology **17** (5), 1506 (2006). |
| 23 | T. Ikehara, R. A. F. Zwijze, and K. Ikeda, Journal of Micromechanics and Microengineering **11** (1), 55 (2001). |
| 24 | A.M. Zaitsev, *Optical Properties of Diamond: A Data Handbook*. (Springer, 2001). |
| 25 | Xinxin Li, Takahito Ono, Yuelin Wang, and Masayoshi Esashi, Applied Physics Letters **83** (15), 3081 (2003); R. I. Pratt, G. C. Johnson, R. T. Howe, and J. C. Chang, presented at the Solid-State Sensors and Actuators, 1991. Digest of Technical Papers, TRANSDUCERS '91., 1991 International Conference on, 1991 (unpublished). |
| 26 | S. D. Bennett, N. Y. Yao, J. Otterbach, P. Zoller, P. Rabl, and M. D. Lukin, Physical Review Letters **110** (15), 156402 (2013). |
| 27 | Linyou Cao, Justin S. White, Joon-Shik Park, Jon A. Schuller, Bruce M. Clemens, and Mark L. Brongersma, Nature MAterials **8** (8), 1476 (2009); Min-Kyo Seo, Jin-Kyu Yang, Kwang-Yong Jeong, Hong-Gyu Park, Fang Qian, Ho-Seok Ee, You-Shin No, and Yong-Hee Lee, Nano Letters **8** (12), 4534 (2008). |
| 28 | Daniel Ramos, Eduardo Gil-Santos, Valerio Pini, Jose M. Llorens, Marta Fernández-Regúlez, Álvaro San Paulo, M. Calleja, and J. Tamayo, Nano Letters **12** (2), 932 (2012). |
| 29 | Parag B. Deotare, Murray W. McCutcheon, Ian W. Frank, Mughees Khan, and Marko LonCar, Applied Physics Letters **94** (12), 121106 (2009). |

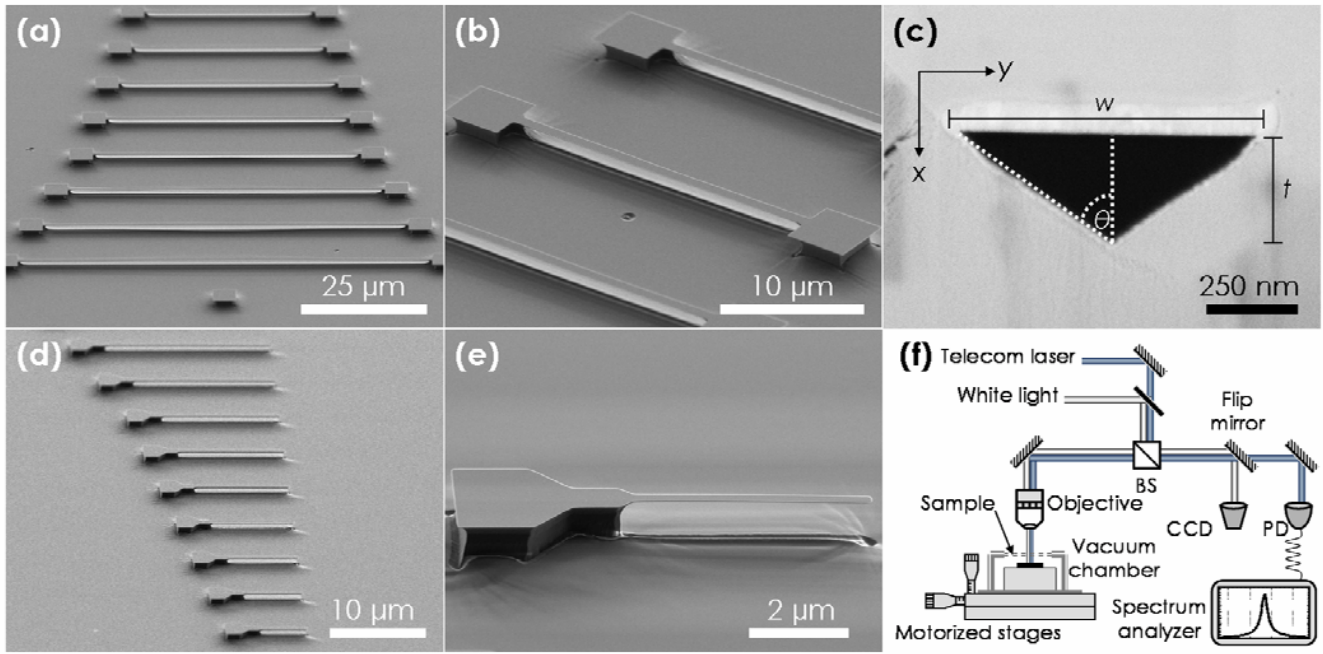

**Figure 1.** **(a)** SEM image of an array of freestanding doubly clamped single-crystal diamond nanobeams. **(b)** SEM image and **(c)** corresponding FIB cross-section of an individual as-fabricated ~ 675 nm wide diamond nanobeam. SEM images of **(d)** an array of freestanding single-crystal diamond cantilevers and **(e)** an individual as-fabricated ~ 675 nm wide diamond cantilever. All SEM images taken at a 60° stage tilt. **(f)** Schematic of the optical characterization setup.

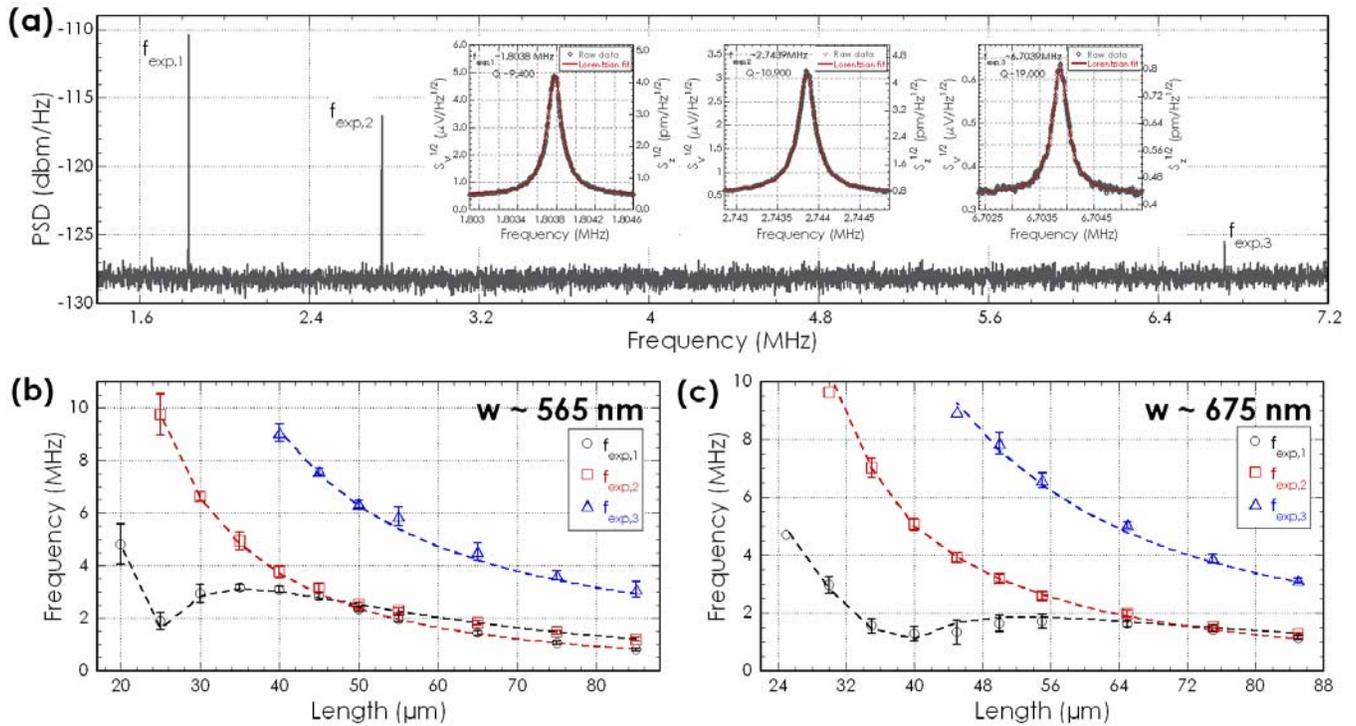

**Figure 2. (a)** Representative thermomechanical power spectral density of a ~ 675 nm wide and 55 µm long doubly clamped diamond nanobeam revealing three resonance peaks, with corresponding high-resolution thermally calibrated noise spectra of the first, second, and third resonance modes shown as insets. Experimentally measured mechanical resonance frequencies of **(b)** ~ 565 nm wide and **(c)** ~ 675 nm wide doubly clamped diamond nanobeams, plotted as a function of nanobeam length. Dashed lines correspond to FEM simulations, with the local minimums in (b) and (c) representing the transition from a compressively-stressed, unbuckled nanobeam to a buckled nanobeam.

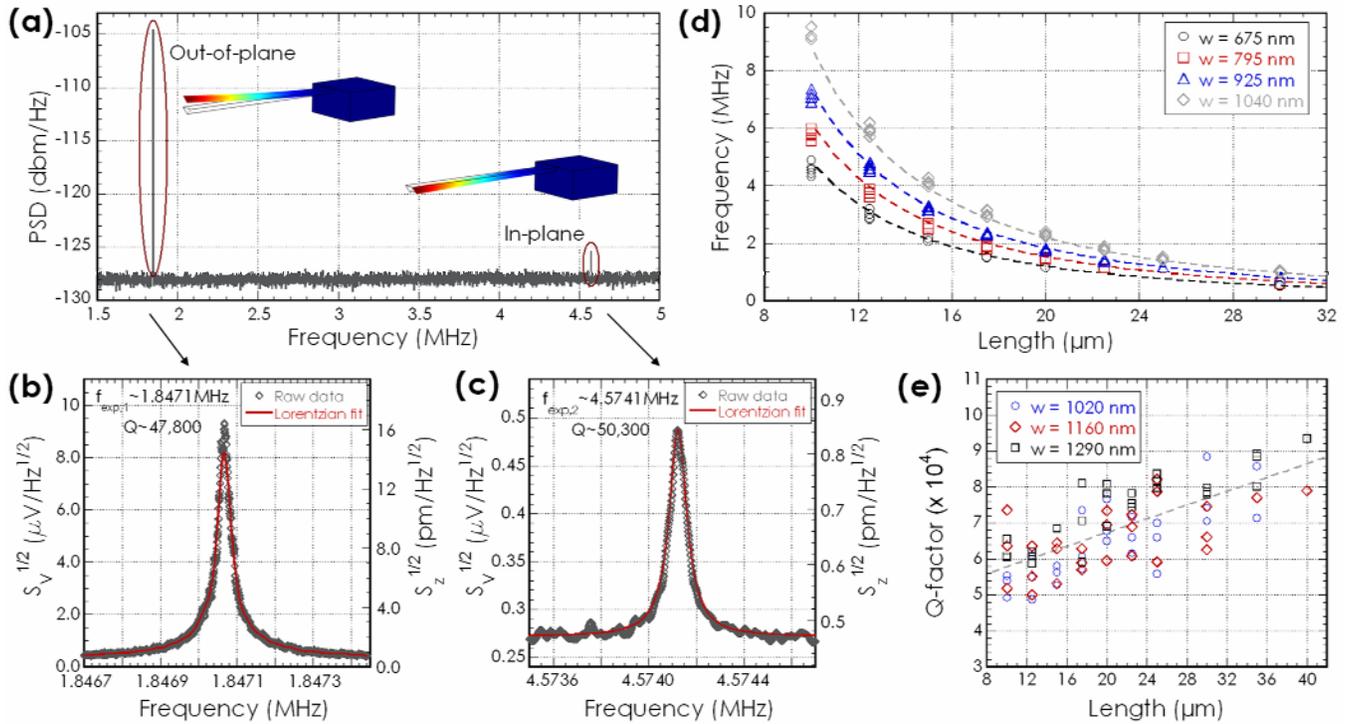

**Figure 3. (a)** Representative thermomechanical power spectral density of an 880 nm wide and 20 μm long diamond nanobeam cantilever revealing out-of-plane (lower frequency) and in-plane (higher frequency) resonance peaks (shown as insets), with corresponding high-resolution thermally calibrated noise spectra of the **(b)** out-of-plane and **(c)** in-plane resonance modes. **(d)** Mechanical resonance frequencies and **(e)** Q-factors estimated from fundamental out-of-plane resonance modes of diamond cantilevers plotted as a function of nanobeam length, with the dashed line given as a guide for the eye.

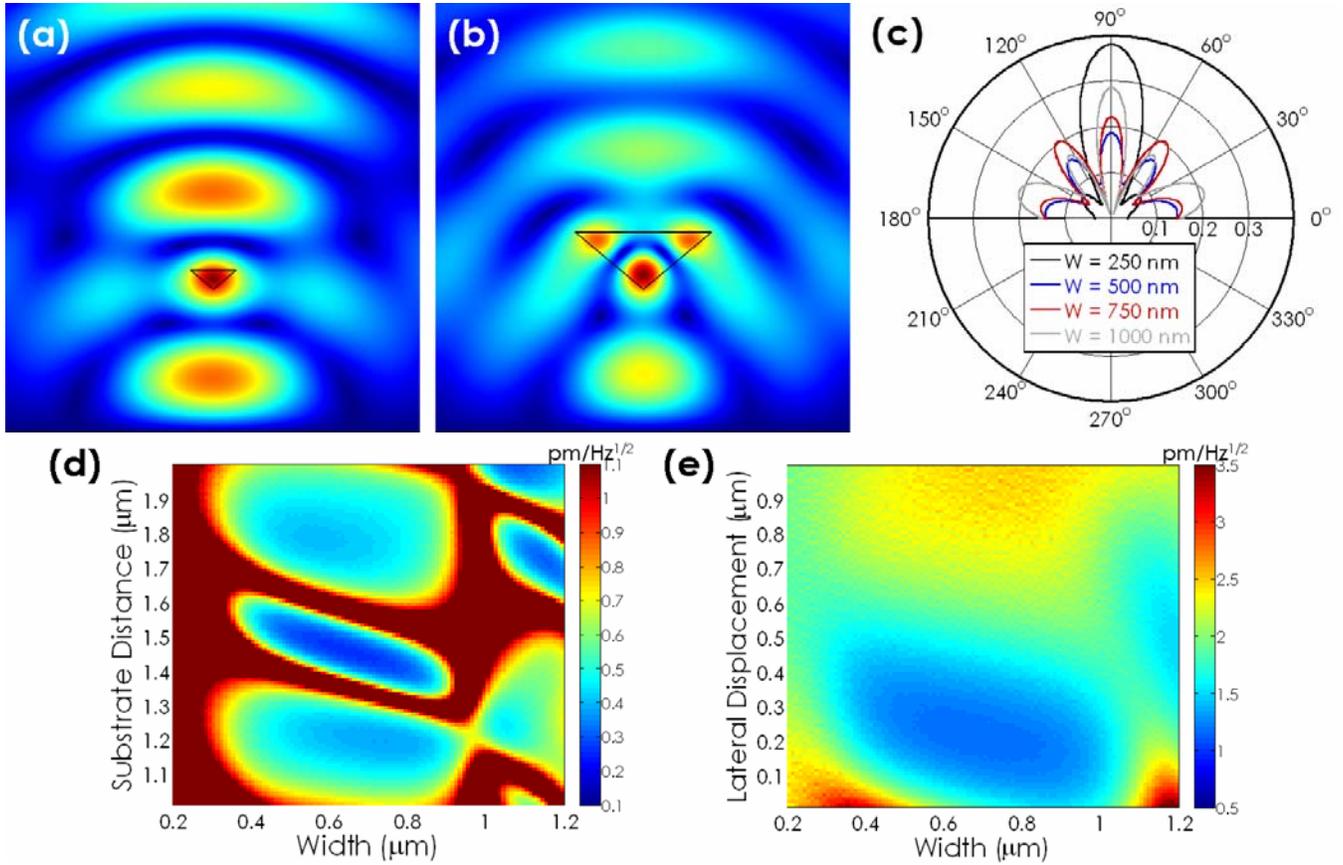

**Figure 4.** Finite element method (FEM) simulated electric field intensity ($\lambda$ = 1500 nm, and 100 μW incident power) spatial confinement by a **(a)** 350 nm and **(b)** 1000 nm wide triangular cross-section diamond nanobeams located 1.0 μm above the substrate. **(c)** Calculated far field profile of electric field intensity reflected by diamond nanobeams of four different widths. Simulated displacement noise spectral density (pm/Hz$^{1/2}$) for **(d)** nanobeam distance above the substrate and **(e)** lateral displacement from the optical axis, as a function of nanobeam width. For the color map in (d), simulated nanobeams are centered on the optical axis. For the color map in (e), the nanobeam substrate distance is fixed at 1.0 μm.